\documentclass[prb,twocolumn,showpacs]{revtex4-1}
\pdfoutput = 1
\usepackage{amsmath}
\usepackage{amssymb}
\usepackage{anysize}
\usepackage{appendix}
\usepackage{graphicx}
\usepackage{dcolumn}
\usepackage{float}

\newcommand{\be}{\begin{equation}}
\newcommand{\ee}{\end{equation}}
\newcommand{\ben}{\begin{equation*}}
\newcommand{\een}{\end{equation*}}
\newcommand{\bea}{\begin{eqnarray}}
\newcommand{\eea}{\end{eqnarray}}
\newcommand{\nn}{\nonumber}
\newcommand{\bean}{\begin{eqnarray*}}
\newcommand{\eean}{\end{eqnarray*}}
\newcommand{\dd}{\mathrm{d}}
\newcommand{\p}{\partial}
\newcommand{\rt}{\right}
\newcommand{\lt}{\left}
\newcommand{\ba}{\begin{array}}
\newcommand{\ea}{\end{array}}

\begin{document}
\title{Modeling Electronic and Thermal Characteristics of Ge/Si-Core/Shell Nanowire Quantum Dot in the Coulomb Blockade Regime}
\bigskip

\author{Shah Mohammad Bahauddin$^1$, Nafiz Ishtiaque$^2$, Ishtiza Ibne Azad$^3$, Zahid Hasan Mahmood$^{1,4}$}
\normalsize
\affiliation{
	$^1$ Dept. of Applied Physics, Electronics \& Communication Engineering, Faculty of Engineering \& Technology, University of Dhaka., Dhaka-1000, Bangladesh \\
	$^2$ Dept. of Theoretical Physics, Faculty of Science, University of Dhaka. \\
	$^3$ Electrical \& Electronic Engineering, Islamic University of Technology, Gazipur, Bangladesh. \\
	$^4$ Semiconductor Technology Research Center, Faculty of Science, University of Dhaka.
}
\date{\today}

\begin{abstract}
	We investigate the transport characteristics of Ge/Si-Core/Shell nanowire with Coulomb Blockade in presence of external magneto-electric fields from a theoretical basis. Using the effective Luttinger-Kohn Hamiltonian we calculate the valence band energy states of the nanowire and find out the state density at mili-Kelvin temperature. We explore the current-voltage and conductance-voltage characteristics. The observed transport characteristics are in close agreement with experimental data. We find these characteristics to be sensitive to the coupling of the quantum dot with the reservoirs. The quantum nature of the characteristics becomes less prominent with increasing temperature, as expected.
\end{abstract}

\pacs{73.23.Hk,73.63.Kv,73.63.-b}

\maketitle

\section{Introduction}
	Engineering spin splitting of charge carriers in semiconductor nanostructures such as quantum wires and dots has opened up realistic opportunities for the production, detection and manipulation of spin currents, allowing coherent transmission of information beyond classical information processing \cite{r1,r2}. It is expected that spatial confinement will have a significant effect on Zeeman splitting when bound-state quantization energies are comparable in nano-devices. Recent advances in fabrication technology have created opportunities to investigate hole spin physics in semiconductor nanowires for a range of different materials \cite{r4,r5}. To achieve the proposal of Loss and DiVincenzo \cite{r6}, scientists thus have fabricated quantum dots integrated with charge sensors namely as quantum point
	contact \cite{r7} to initialize, manipulate and readout electron spins in GaAs systems \cite{r8}. However, the attempt suffers from difficulties in control and lacking of spin coherence due to the overlap of the electronic wavefunctions in these systems with a large number of nuclear spins \cite{r9}. Though this intrinsic decoherence rate has been successfully reduced by spin-echo technique \cite{r10}, the complexity of the system remains as an issue under debate.
	
	As a result, the prospect of long coherence times in group IV materials due to the predominance of spin-zero nuclei \cite{r11} has stimulated several proposals \cite{r12} and significant experimental effort. Experimental progress includes realizations of DQDs in carbon nanotubes \cite{r13} and Si:P \cite{r14} as well as single dots in Si and Ge/Si nanowires \cite{r15} and Si/Ge 2DEGs \cite{r16}. Moreover, the observation of long relaxation times due to negligible interaction with nuclear spin bath makes the spin of a hole in a semiconductor quantum dot the best contender for solid-state qubit. Thus, Ge/Si-Core/Shell nanowire has become a central attraction for both the experimental and theoretical investigation since the valence band offset at core-shell interface provides a strong radial confinement of holes and due to the large sub-band spacing, this behaves as one-dimensional hole gas at ultra low temperature \cite{r17}.
	
	In this work, we explore the Coulomb Blockade effect in current-voltage and conductance-voltage characteristics of Ge/Si-Core/Shell nanowire Quantum Dot under perpendicular electric and magnetic fields. The Valence band states of the nanowire are theoretically analyzed using Luttinger-Kohn Hamiltonian \cite{r23, r24}. Then a self-consistent field approximation using non-equilibrium Green’s Function formalism is performed to simulate density of states and from there we investigate the transport characteristics in mili-Kelvin temperatures.

\section{Model Hamiltonian and Self-Consistent Solution}
	In our present analysis we use an effective 1D Luttinger-Kohn Hamiltonian for a cylindrical nanowire with Ge core of radius, $R = 5$ nm and Si shell of thickness $t = 2$ nm. This Hamiltonian is perturbative and corresponds to ground state and first excited state only. The electrochemical potentials of the materials are comparable to the energy of these states and the applied potentials are in mili-volt order, hence the higher excited states will not contribute significantly in current conduction. For our calculation we assume the confinement to be an infinite cylindrical well, since the valence band offsets in Si/Ge Core/Shell nanowire is large. Now, the four eigenstates, $|g_\pm\rangle$ and $|e_\pm\rangle$, corresponding to ground and excited states for
	$F_z = \pm 1/2$ ($F_z$ is quantum number such that, $[H_{LK} + V, F_z] = 0$ and the operator $F_z = L_z + J_z$, $L_z$ being orbital angular momentum along the wire axis) at ݇$k_z = 0$ will serve as the basis states for the Hamiltonian. Here, we have added the contribution of electric field that causes Rashba Spin Orbit Interaction \cite{r28}, the effect of external magnetic field which lifts the Kramer degeneracy and the spin splitting due to strain which is induced by the compression of Ge core by Si shell. A detailed derivation of the Hamiltonian for Ge/Si core/shell nanowires is provided elsewhere \cite{r18,r26}; here we quote the matrix needed to calculate hole spectrum,
	\be
		H_{NW} = \lt( \begin{array}{cccc}
		H_{11} & H_{12} & H_{13} & H_{14} \\
		H_{12}^* & H_{22} & H_{23} & H_{24} \\
		H_{13}^* & H_{23}^* & H_{33} & H_{34} \\
		H_{14}^* & H_{24}^* & H_{34}^* & H_{44}
		\end{array}\rt)
	\ee
	where,
	\ben\begin{split}
		H_{11} &= H_{33} = \frac{\hbar^2k_z^2}{2m_g} \\
		H_{22} &= H_{44} = \frac{\hbar^2k_z^2}{2m_e} + \frac{0.73\hbar^2}{m_0R^2} + \delta(\gamma) \\
		H_{12} &= H_{34} = -i\frac{7.26\hbar^2}{m_0R}k_z \\
		H_{13} & = \mu_BB_x(X_1+X_2);\; H_{24} = \mu_BB_x(X_1-X_2) \\
		H_{14} &= -H_{23} = -i\mu_BB_xX_3 - ieE_yU_0
	\end{split}\een
	with, $B_x = 10$ T, $E_y = 1$ V/m. For Ge, $U_0 = 0.15R$, $X_3 = 8.04R$. $X_1$ and $X_2$ are material dependent constants and $\delta(\gamma)$ rescales the energy splitting and depends on the relative shell thickness \cite{r26}, $\gamma = t/R$. The energy eigenvalues are: -4.675 meV, -1.45 meV, 5.075 meV, and 8.15 meV.

\section{Density of states}
	We denote the creation and annihilation operators of the ground state and the first excited state by $c, c^+$ and $d, d^+$ respectively. The Schr\"{o}dinger equation for these operators incorporating the Coulomb interaction between them becomes:
	\be i\hbar\frac{\dd}{\dd t} c = \epsilon c + \sum_r \tau_r c_r + Ud^+dc \label{e1}\ee
	Here $\epsilon$ is the energy of the energy level $c$, $\tau_r$ is the coupling between the level $c$ of the channel and the $r$-th energy level of the reservoir, and $\epsilon_r$ is the energy of the $r$-th reservoir energy level. $U$ is the coulomb potential due to interaction between the two energy levels of the channel. We write the uncoupled reservoir operators as $C_r$ and define a source term $S^C := \sum_r C_r$ in Fourier space for the channel operators. Then the Fourier space solution of (\ref{e1}) can be written as $\tilde c = G(E) S^c$ where $G(E)$ is the Green's function \cite{r25}:
	\be G(E) = \frac{1-\langle d^+d \rangle}{E-\epsilon-\Sigma_0} + \frac{\langle d^+d \rangle}{E-\epsilon-U-\Sigma_0} \label{e2}\ee
	with $\Sigma_0$ given by \be \Sigma_0 = \sum_r \frac{|\tau_r|^2}{E-\epsilon_r+i\eta}. \ee
	The density of states, denoted by $D(E)$, is calculated using self-consistent approximation using the Green's function (\ref{e2}). We find, \be D(E) = i(G-G^*). \label{e3}\ee The external magnetic field breaks the degeneracy of both the ground state and the first excited state and each of these states separates into two different states corresponding to spin up and down.

\section{$I$-$V$ characteristics}
	We assume that the temperature is low enough for particle production due to thermal excitation to be ignored. Therefore amount of charge in the channel at any time is well defined. We also assume a constant capacitance as in the constant interaction model \cite{r29}. From the density of state (\ref{e3}) the current $I$ is easily evaluated,
	\bea && I(V_G, V_{SD}) = \frac{e\,\gamma_1\gamma_2}{\hbar(\gamma_1+\gamma_2)} \nn\\ &\times& \int_\infty^\infty \dd E\, D(E-U-V_G) (f_S(E+V_{SD}/2) \nn\\ &-& f_D(E-V_{SD}/2)) \label{e4}\eea
	where $f_S$ and $f_D$ are Fermi functions of the source and drain.
	
\begin{figure}[ht]
	\begin{center}\leavevmode
		$$\begin{array}{c}
			\includegraphics[scale=.46]{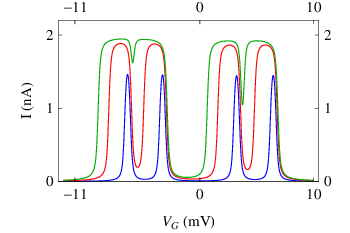} \\
			\includegraphics[scale=.46]{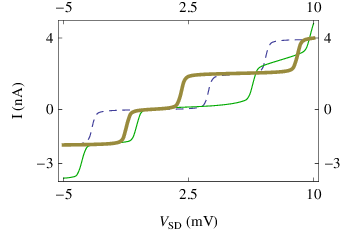}
		\end{array}$$
		\caption{(Top) Variation of current with respect to gate voltage. Here, Blue: $V_{SD} = 0.5$ mV, Red: $V_{SD} = 1.5$ mV, Green: $V_{SD} = 3$ mV. (Bottom) Variation of current with respect to drain-to-source voltage. Here, dashed: $V_G = 0$ mV, Green: $V_G = -2.5$ mV, Brown: $V_G = -5$ mV.}
		\label{IV500mK}
	\end{center}
	\end{figure}

	First we observe the variation of current with varying gate voltage ($V_G$) at constant source-drain voltage ($V_{SD}$). In Fig. \ref{IV500mK} (top) discrete current peak appears, meaning Coulomb blockade is lifted by changing the gate voltage. When the electrochemical potential of the dot in presence of one hole, is in the bias window, in our case $(\mu_S, \mu_D)=($0.2 meV, 0.3 meV), one hole tunnels into the dot from the source, so that the number of holes increases from 0 to 1, causing the first peak at -6.4 mV. This single hole tunneling continues and the electrochemical potential level corresponding to transport rises
	sequentially and the number of holes increases upto 4 for $V_G = -3.25$ mV, 3.2 mV and 6.5 mV. In the valleys between these peaks, the numbers of holes on the dot is fixed due to Coulomb Blockade. By tuning the gate voltage from one valley to the next one, the number of holes on the dot thus can be precisely controlled. The distance between the peaks corresponding to $E_{\mathrm{add}}$ which in our case are found 3.15 meV for ground spin states and 3.3 meV for excited spin states respectively.
	
	We now look at the regime where $V_{SD}$ is so high that multiple dot levels can participate in hole tunneling. A positive $V_{SD}$ increases the difference of electrochemical potentials of the source and the drain. Due to capacitive coupling the levels of the dot are also increased. In our case we assume, $C_G = 3$ aF, $C_D = C_S = 10$ aF. Thus, for $V_{SD} = 1.5$ mV, the hole states are separated further and Fig. \ref{IV500mK} (top) shows that the number of holes increases from 0 to 1 at -7.1 mV, 1 to 2 at -3.85 mV, 2 to 3
	at 2.5 mV and 3 to 4 at 5.9 mV respectively. The distance between the peaks corresponding to $E_{\mathrm{add}}$ which are found to be 3.25 meV for ground spin states and 3.4 meV for excited spin states respectively. Moreover, increasing $V_{SD}$ signifies transition involving excited state falls within the bias window and double hole tunneling occurs which implies the number of holes alternates between 0, 1 and 2 leading to a double hole tunneling. According to Fig. \ref{IV500mK} (top), it can be concurred that the peak current increases from approximately 2.35nA to 3.18nA for the variation of source to drain voltage from 0.5 mV to 3 mV.
	
	In Fig. \ref{IV500mK} (bottom) when the voltage is sufficient for a states of $N = 1$ hole to be transferred to the dot to the reservoir, a sudden jump of the current will first be observed due to the junction with a high tunnel rate. Then the current will grow slowly, because of the low-rate junction coming into play, until states of $N = 2$ hole appears on the dot.  Thus, although the current through the system passes continuously, a particular voltage dependent amount of holes will exist showing "Coulomb staircase".
	
	\begin{figure}[ht]
	\begin{center}\leavevmode
		\includegraphics[scale=.46]{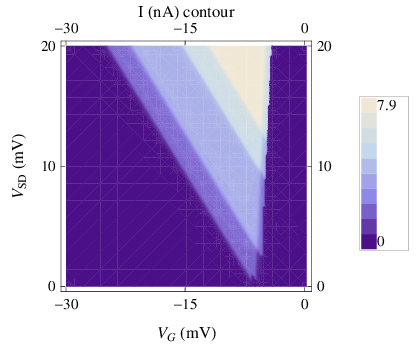}
		\caption{Contour plot of $I$-$V_G$-$V_{SD}$. Here, transport is in $z$-axis, radius of Ge-core $R = 5$ nm, Rashba SOI $E_y = 1$ V/m, applied magnetic field, $B_x = 10$ T, temperature = 500 mK.}
		\label{IVV500mK}
	\end{center}
	\end{figure}
	In Fig. \ref{IVV500mK}, we plot a contour of current in the $V_G$-$V_{SD}$ plane as a representation of the $I$-$V$ characteristics. Here, each time a hole is added to the quantum dot, a current peak appears as a diagonal line, a typical signature of a single quantum dot \cite{r27}.

\section{Conductance characteristics}
		Now from the expression of current (\ref{e4}) we evaluate the conductance, \be \sigma(V_G, V_{SD}) = \frac{\p}{\p V_{SD}} I(V_G, V_{SD}). \ee The conductance response of the device completely characterizes it and here we observe the contours of conductance in the $V_G$-$V_{SD}$ plane.
		\begin{figure}[ht]
		\begin{center}\leavevmode
			\includegraphics[scale=.46]{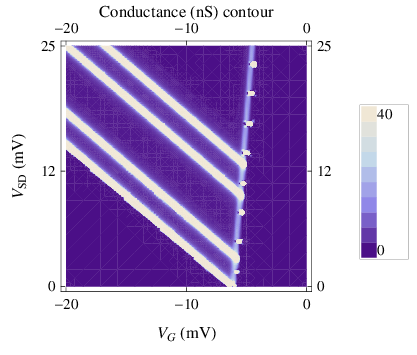}
			\caption{Contour plot of conductance-$V_G$-$V_{SD}$. Here, transport is in z-axis, radius of Ge-core R = 5 nm, Rashba SOI $E_y = 1$ V/m, applied magnetic field, $B_x = 10$ T, temperature = 500 mK.}
			\label{cond}
		\end{center}
		\end{figure}
		
		To understand Fig. \ref{cond}, first the electrochemical potential ladder is defined by the gate voltage axis of the ($V_{SD}$ and $V_G$) plot. Now, by sweeping the gate voltage at low bias regime, hole tunneling occurrs due to the transition $|g_-\rangle \leftrightarrow |g_+\rangle$. For all other gate voltages the dot is in Coulomb blockade. The figure also depicts several sets of bright lines. These are the regions of maximum differential conductance, i.e., the dot current changes in these sets of $V_{SD}$ and $V_G$. The slopes of the V-shaped lines at 6  mV depend on the assumed capacitance and thus defines the regions of Coulomb Blockade. The other solid lines indicate where the current changes due to the transitions involving excited states. From Fig. \ref{cond}, it is evident that the separations of the ground and excited states for a single hole are 3 mV, 9.5 mV, and 12.75 mV.

\section{Effects of temperature}
	\begin{figure}[ht]
	\begin{center}\leavevmode
		$$\begin{array}{c}
			\includegraphics[scale=.46,trim=0 45 0 0,clip]{IVg500mK.png} \\
			\includegraphics[scale=.46,trim=0 0 0 17,clip]{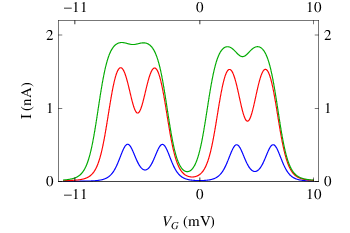}
		\end{array}$$
		\caption{Variation of current with respect to gate voltage for $ T = 500$ mK (top) and 5.0 K (bottom). Here, Blue: $V_{SD} = 0.5$ mV, Red: $V_{SD} = 1.5$ mV, Green: $V_{SD} = 3$ mV.}
		\label{IVDST}
	%\end{center}
	%\end{figure}
	%\begin{figure}[ht]
	%\begin{center}\leavevmode
		$$\begin{array}{c}
			\includegraphics[scale=.46,trim=0 45 0 0,clip]{IVds500mK.png} \\
			\;\includegraphics[scale=.46,trim=0 0 0 17,clip]{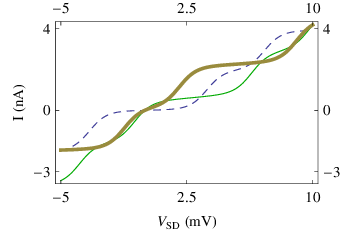}
		\end{array}$$
		\caption{Variation of current with respect to drain-to-source voltage for $ T = 500$ mK (top) and 5.0 K (bottom). Here, dashed: $V_G = 0$ mV, Green: $V_G = -2.5$ mV, Brown: $V_G = -5$ mV.}
		\label{IVGT}
	\end{center}
	\end{figure}
	In this section, we present a comparative study of the temperature dependence of hole transport. We simulate the self-consistent charge model in coulomb blockade regime in two different temperatures, 500 mK and 5.0 K, and obtain the corresponding $I$-$V$ curves and contours of conductance-voltage. With increasing temperature, it is evident that the transport should become more classical and hence quantum behavior of the system should be less prominent.
	
	In Fig. \ref{IVDST}, one can see that the peaks of current with respect to gate voltage are smearing out with increase in temperature. This is because energy levels become excited due to thermal fluctuations and thus discreteness of the curve tends to vanish.
	\begin{figure}[ht]
	\begin{center}\leavevmode
		$$\begin{array}{cc}
			\includegraphics[scale=.47,trim=0 23 0 0,clip]{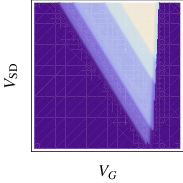} &
			\includegraphics[scale=.47,trim=31 23 0 0,clip]{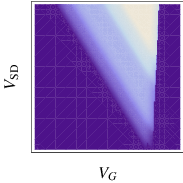} \\
			\includegraphics[scale=.47,trim=0 0 0 0,clip]{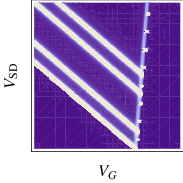} &
			\includegraphics[scale=.47,trim=31 0 0 0,clip]{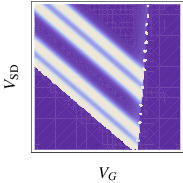}
		\end{array}$$
		\caption{Contour plot of $I$-$V_G$-$V_{SD}$ (top) and conductance-$V_G$-$V_{SD}$ (bottom) for $ T = 500$ mK (left) and 5.0 K (right). Here, transport is in z-axis, radius of Ge-core $R = 5$ nm, Rashba SOI $E_y = 1$ V/m, applied magnetic field, $B_x = 10$ T.}
		\label{CondT}
	\end{center}
	\end{figure}
	In Fig. \ref{IVGT}, we have plotted current with respect to drain-to-source voltage. Due to thermal excitation, the staircase tends to be linear and becomes dispersed with increasing temperature.
	
	Finally, in Fig. \ref{CondT}, contours of current $I$ and conductance $\sigma$ in the $V_{G}$-$V_{SD}$ plane are plotted and they also lose their discrete nature with the rise in temperature from 500 mK to 5.0 K. All these results are physically consistent and can be explained by thermal excitation.
	\vspace{.9cm}

\section{Discussion}
	In our analysis we have been able to reproduce most the prominent features of the Ge/Si-Core/Shell nanowire as found in experiments \cite{r17,r19,r20} and the developed model is capable of describing most of the detailed features by taking into account the properties of the materials. The self-consistent field approximation using non-equilibrium Green’s function formalism simulates realistic density of states and transport characteristics in mili-Kelvin temperatures. We observe the quantization of $I$-$V_G$ and the staircase profile of $I$-$V_{SD}$. Additionally, the contour of Conductance-$V_G$-$V_{SD}$ characterizes the device operating in single hole transition. We find these characteristics to be significantly sensitive to
	the coupling of the dot to the reservoir by means of capacitance effect. They are also sensitive to the operating temperature and start losing their quantum nature with rise in temperature. Hence, from a theoretical point of view, this work suggests that Ge/Si-Core/Shell nanowire is an attractive candidate for quantum information processing. However, several directions seem promising for further explorations. For this particular device structure, even cleaner wires would show evidence of asymmetric geometry \cite{r21,r22} and should be accounted for in theoretical model. Moreover, soft wall consideration along with wave function penetration can be added to achieve results with better accuracy.

\end{document}